 \documentclass[twocolumn]{emulateapj}

\def\Ha{H$\alpha$}
\def\kms{km$\;$s$^{-1}$}
\def\Rs{R$_\odot$}

\slugcomment{to be submitted to PASP}

\shorttitle{Rapid H$\alpha$ variability in T CrB}
\shortauthors{Zamanov et al.}

\begin{document}


\title{Rapid H$\alpha$ variability in T Coronae Borealis \altaffilmark{*}}


\author{R.Zamanov$^{1,2}$, A. Gomboc$^{1,3}$, M. F. Bode$^1$, J.M. Porter$^1$, and N.A. Tomov$^{2}$}
\affil{$^1$Astrophysics Research Institute, Liverpool John Moores University, 
              Twelve Quays House, Birkenhead, CH41 1LD, UK \\
	      $^2$National Astronomical Observatory Rozhen, POB 136, Smoljan, BG-4700, Bulgaria \\
	      $^3$Faculty of Mathematics and Physics, University in Ljubljana, Jadranska 19, 1000 Ljubljana, Slovenia}

\altaffiltext{*}{based on observations obtained in 
National Astronomical Observatory Rozhen, Bulgaria}
\altaffiltext{3}{e-mail:  rz@astro.livjm.ac.uk; ag@astro.livjm.ac.uk; mfb@astro.livjm.ac.uk;  jmp@astro.livjm.ac.uk}



\begin{abstract}
   We report our search for variability in the \Ha\ emission line  of the 
   recurrent novae T CrB 
   with time resolution  10-15 minutes. This is comparable with the time scale 
   of the photometric flickering observed in this object. This is the first time that
   observations of the short time scale variation in emission lines
   have been made in 
   this object. 
   On two nights (990106 and 990107) we detected statistically significant 
   variability (at 0.99 confidence level) in the line profile of \Ha. 
   This variability is confined in the central part  
   of the emission line ($\pm$100 \kms)
   although  FWZI(\Ha) is $\sim$800 \kms.
   The variability in the line profile is accompanied 
   with  variability of the total equivalent width, EW(\Ha), $\pm$8\% for 990106, 
   and $\pm$6\%  for 990107 (calculated from the mean value of EW).
   Assuming Keplerian motion, the variability is 
   generated at a distance $\sim$20-30\Rs\ from the white dwarf, 
   which is approximately the radius of the ring the stream of gas 
   forms as it flows away from L$_1$. 
   On three other nights  we are only able to put upper limits on the variability, 
   $\Delta$EW(\Ha)
   $\pm$2\% for 980415, 
   $\pm$4\% for 980802, and 
   $\pm$3\% for 980803.   
\end{abstract}





\section{Introduction}

  The short term photometric behaviour of symbiotic stars has been investigated 
  in many systems (e.g. Dobrzycka et al., 1996;  Sokoloski et al, 2001).
  However, there are no systematic searches for rapid {\it spectral} variations.
  Until now the search for rapid spectral variability has been undertaken 
  in only three symbiotics:  CH~Cyg, MWC~560,  and RS~Oph. 
  All three stars are known to exhibit flickering and have collimated  outflows.
  Rapid spectral variability in the Balmer lines of CH~Cyg 
  on time scales of $\sim$1 hour have been detected and is probably
  connected with blobs ejected from a white dwarf,   
  acting as a propeller (Tomov et al. 1996). 
  For RS~Oph,  Sokoloski (2002) reported variability on time scales of hours 
  in the HeII$\lambda$4686 line although  no variability in H$\beta$ was detected.
  No significant variability has been detected in MWC~560 (Tomov et al. 1995).
 
  In a few cataclysmic variables (CVs), which are closely related to
  symbiotic stars, rapid spectral changes are visible 
  in optical and UV lines as a result of a variable accretion disk wind:
  BZ Cam (Ringwald \& Naylor 1998),
  V592 Cas (Witherick et al. 2003), 
  RW Sex (Prinja et al. 2003),
  V~603 Aql (Prinja et al. 2000a),
  BZ Cam (Prinja et al. 2000b).
  
  The detection of variability on time scales of tens of minutes in 
  symbiotic stars and cataclysmic variables 
  provides the motivation for this study.
  T CrB consists of a red giant and a hot component, most probably a white dwarf. 
  T CrB can  be classified as a cataclysmic variable, symbiotic star, and recurrent nova 
  and accretes at a rate 1$\times$10$^{-8}$M$_\odot$ yr$^{-1}$ 
  (Selvelli et al. 1992). This places it 
  among the cataclysmic variables with the highest accretion rates
  (see Warner 1995, p.476) and for such high accretion rates 
  an accretion disk wind could be expected.  
  Here we report our search for signatures of variability 
  in time-resolved optical spectroscopy in the \Ha\ line of  T~CrB,
  with time resolution $\sim$10-15 minutes. This resolution is comparable 
  with the time scale of flickering (see for example Zamanov et al. 2004).

\section{Observations}

The observations were performed  
with the Coud\'e spectrograph of the 2.0 m RCC telescope 
at the Bulgarian National Astronomical Observatory ``Rozhen''.
The spectra have a dispersion of 0.2 \AA\ pixel$^{-1}$. 
The normal CCD reduction procedures 
(i.e. bias subtraction, flat fielding, wavelength calibration, etc) 
have been undertaken in the IRAF environment. A 
journal of observations is given in Table~1.
The  signal to noise ratio (SNR) achieved 
on the individual exposures is 35-65. 

\begin{figure}
   \centering
   \includegraphics[width=9cm]{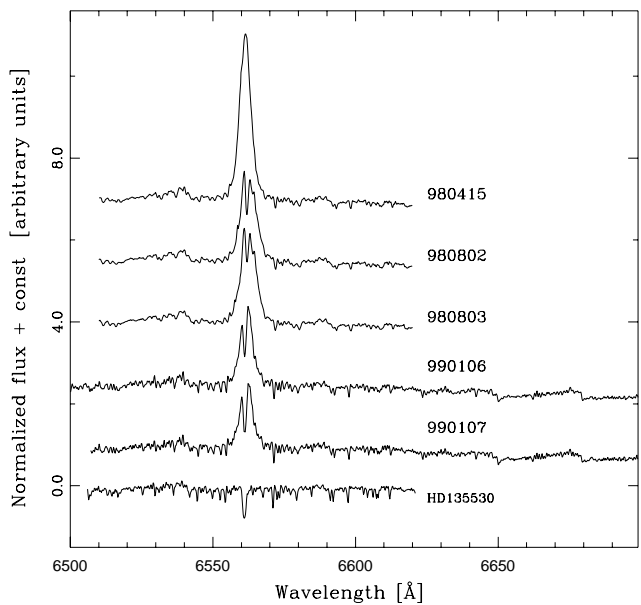}
      \caption{The nightly average spectra for  all five nights. 
      A spectrum of the red giant  HD135530 (M2III) is shown for comparison at the
      bottom. }
         \label{FigAvBroad}
\end{figure}

The averaged spectra, normalised to the local continuum, are presented in
Fig.\ref{FigAvBroad}, along with a comparison spectrum of the red
giant HD135530.

 \begin{centering}
 \begin{table*}
 \caption{\Ha\  observations of T CrB. In the table are given the date of observations,
 the beginning  and the end of spectroscopic observations and truncated Julian Day.
 Orbital phase is calculated using the ephemeris of Fekel et al. (2000). 
 The following columns give wavelength coverage, number of exposures in  each night and
 the exposure time of the individual exposures.  
 For the EW(\Ha) we give the mean, standard deviation, plus the smallest and the
 largest values for each night.}
  \begin{tabular}{lccccrrr}
  Date    & UT$_{start}$ - UT$_{end}$ & TJD	   & phase  & wavelength   & N x Exp.time &  EW(\Ha)        & EW$_{min}$ -- EW$_{max}$ \\
  yymmdd  & hh:mm - hh:mm	      & mid	   &	    & coverage     &              &  [\AA]          & [\AA] 	       \\
          &			      & 	   &	    &  [\AA]	   &		  &                 &   	       \\
          &			      & 	   &	    &		   &		  &                 &  	               \\
  980415  & 22:37 - 01:43	      & 50919.512  & 0.187  & 6510 - 6620  & 10 x 15 min  & 20.84 $\pm$ 0.31& 20.39 -- 21.36   \\
  980802  & 18:50 - 20:58	      & 51028.328  & 0.665  & 6510 - 6620  & 12 x 10 min  & 13.39 $\pm$ 0.23& 12.72 -- 13.81   \\
  980803  & 18:23 - 21:03	      & 51029.305  & 0.669  & 6510 - 6620  & 15 x 10 min  & 13.97 $\pm$ 0.25& 13.63 -- 14.45   \\ 
  990106  & 02:59 - 04:13	      & 51185.652  & 0.356  & 6500 - 6700  &  7 x 10 min  &  9.77 $\pm$ 0.62&  8.90 -- 10.51   \\
  990107  & 01:50 - 02:51	      & 51186.598  & 0.360  & 6507 - 6707  &  3 x 20 min  &  7.92 $\pm$ 0.45&  7.45 --  8.36   \\
          &			      & 	   &	    &		   &		  &                 &  	        \\	 
 \label{tab1}									   				    
 \end{tabular}									   				    
 \end{table*}	
 \end{centering} 


\section{Results}

\subsection{Average, fractional and temporal variance spectra}

The prominent emission line in Fig. \ref{FigAvBroad} is \Ha.
On a time scale of years the intensity of the \Ha\ line varies
from EW$=$0.5 \AA\ up to 35 \AA\  (Iijima 1990, Anupama \& Prabhu 1991, 
Zamanov \& Mart\'{i} 2000, Stanishev et al. 2004).
During the search for rapid variability 
we have observed a range of different profiles. 
An atlas of the different \Ha\ profiles  is given in Stanishev et al. (2004).
In the spectra obtained on 15th April 1998 the line is single peaked, 
and the central absorption dip is not visible.
During August 1998, the profile is a little unusual, with 3 peaks or 2 absorption dips.
During January 1999, the profile exhibits double peaks with a well 
defined central dip, which is partly due to the \Ha\ absorption line of the red giant.

\begin{figure}
   \centering
   \includegraphics[width=9cm]{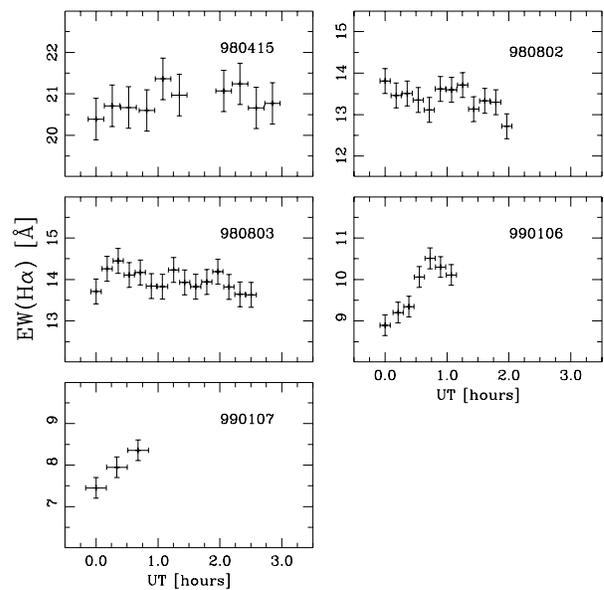}
      \caption{Measurements of the EW(\Ha). 
      The vertical error bars indicate the errors in EW measurements, 
      the horizontal the exposure time of the spectra.
      In the first three nights the variability is of the order 
      of the estimated errors.
      On 990106 and 990106 it is more or about 5 times the error of the measurements.
            }
         \label{EWHa}
 \end{figure}

The total  EW(\Ha) was measured on each exposure in our sample. 
The results are presented in Fig.\ref{EWHa}. 
In the first three nights the variability is less than three times the estimated errors.
On 990106 and 990107 it is about 5 times the error of the individual 
measurements.

The measurements of the EW depend strongly on the 
continuum placement. The continuum was defined as a stright line 
or spline fitting, using parts of the spectrum located 
on both sides of \Ha.
We used different wavelength intervals to produce different normalisations.
This leads to changes of the absolute value of EW, but keeps the relative 
distributions of the points the same. The errors of the EW have been
calculated from four  measurements, with different normalizations.

In order to search for the variability in the line profile 
of H$\alpha$ line  during an individual night, 
we calculate the fractional variance spectrum based on all the data  
in each  night:
\begin{equation}
\sigma = \sqrt{ {{{1\over {(N-1)\bar{f}^2_\lambda}} 
{\sum_{i=1}^N {(f_{i\lambda}-\bar{f}_\lambda)^2}}}}} ,
\end{equation}
where $N$ is the number of spectra, $f_{i\lambda}$ is the flux at wavelength 
$\lambda$ of the $i^{th}$ 
spectrum and $\bar{f}_\lambda$ is the flux at that wavelength averaged over all spectra.
The $\sigma$ in the continuum should correspond to (SNR)$^{-1}$, and hence a peak in 
$\sigma$ should  determine the significance of a variation.

One other statistical technique which is particularly useful to
accurately determine the spectral location of variability is 
the temporal variance spectrum 
(Prinja et al. 2003;  Fullerton, Gies \& Bolton 1996): 

\begin{equation}
(TVS)_\lambda = {s_0^2 \over {(N-1)}} 
{\sum_{i=1}^N { {(f_{i\lambda}-\bar{f}_{w\lambda})^2 } \over {s_{i}^2 f_{i\lambda} } } }
\end{equation}
where  $\bar{f}_{w\lambda}$ is the weighted mean of the normalised intensity, 
s$_{i}$ is the inverse of the SNR of spectrum $i$ measured in the continuum,
and 
$s_0^2 = [1/N  \sum_{i=1}^N {s_{i}^{-2}}  ]^{-1}$.

\begin{figure*}[htb]
 \mbox{}
 \vspace{17.0cm}
  \includegraphics{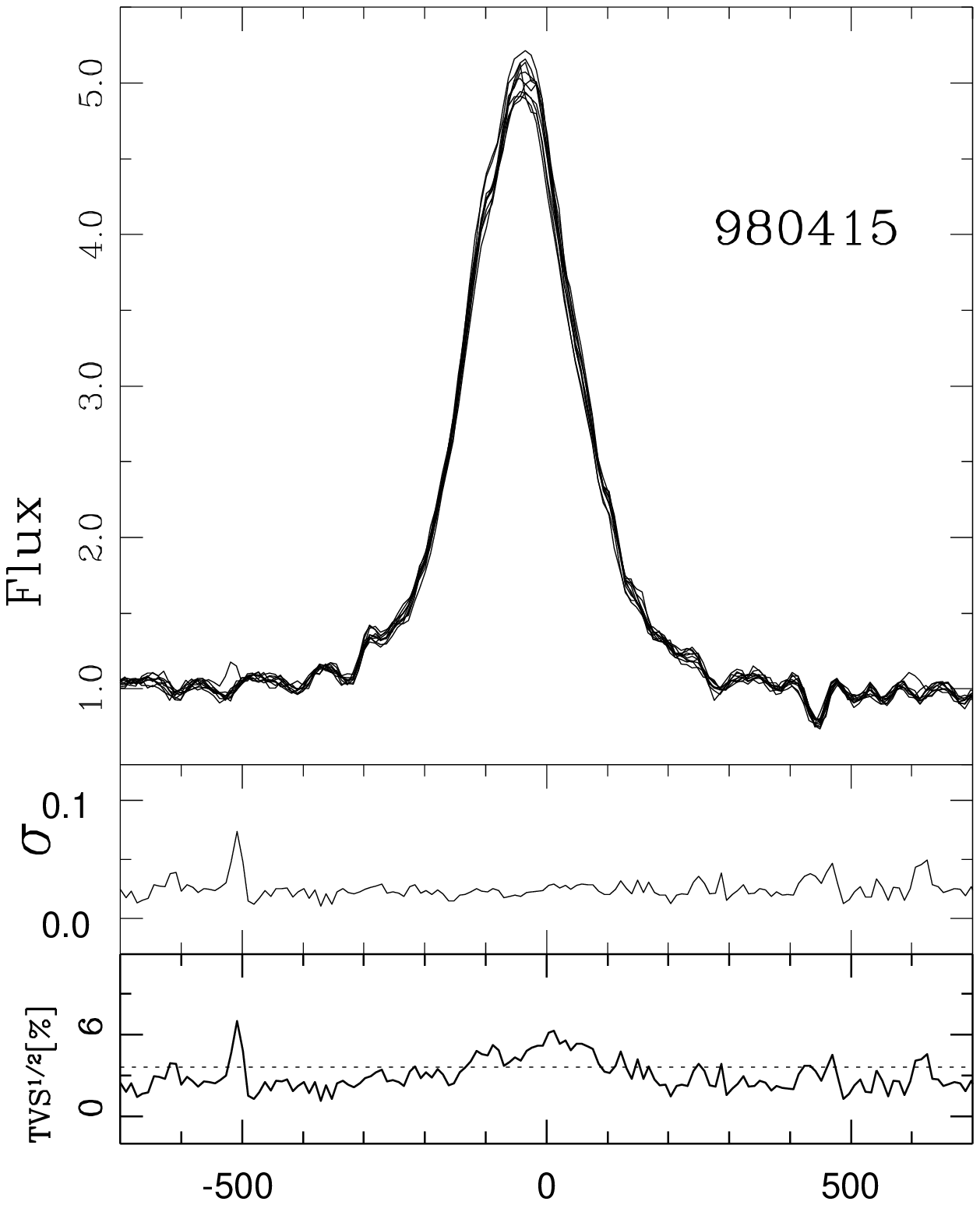}
  \includegraphics{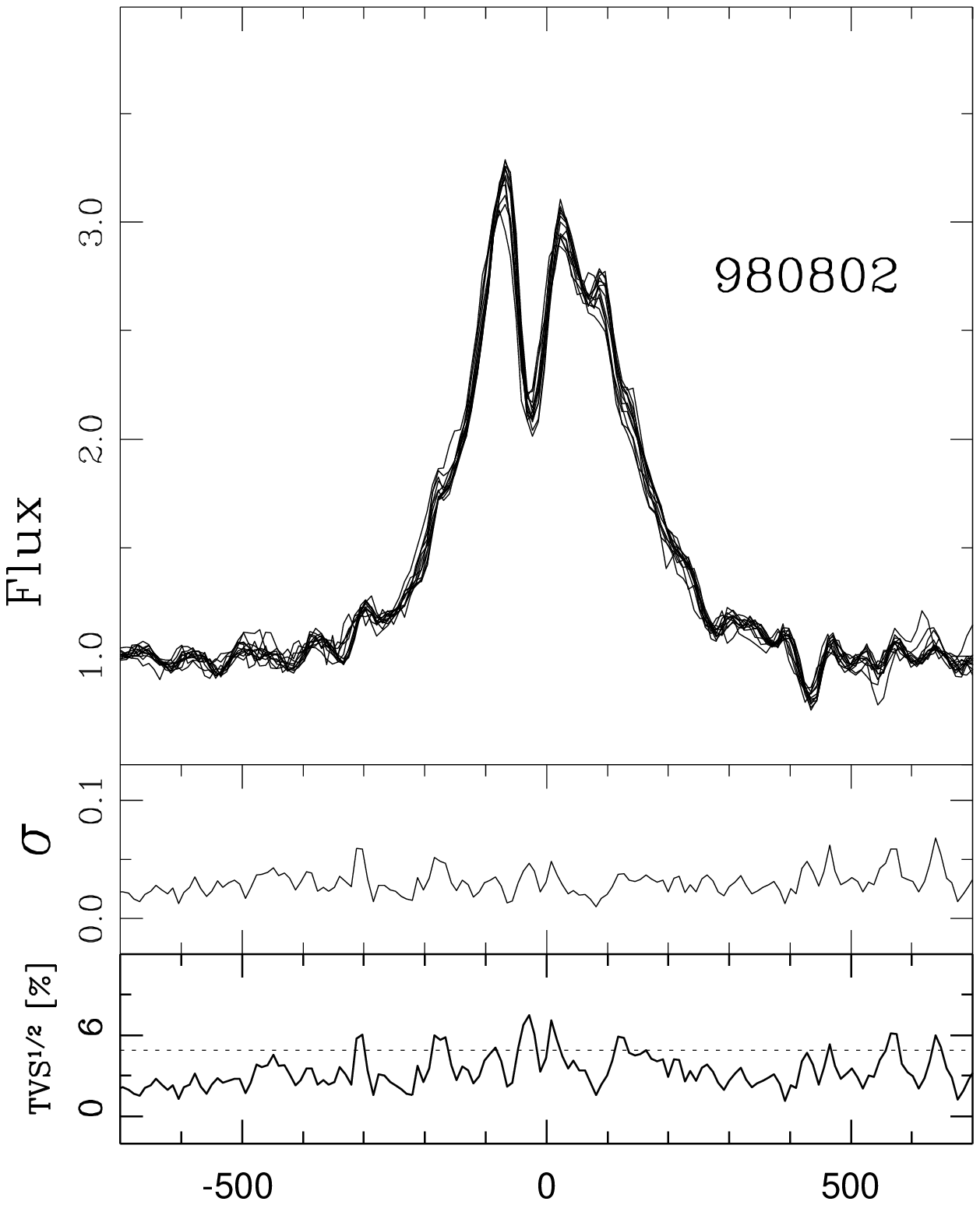}
  \includegraphics{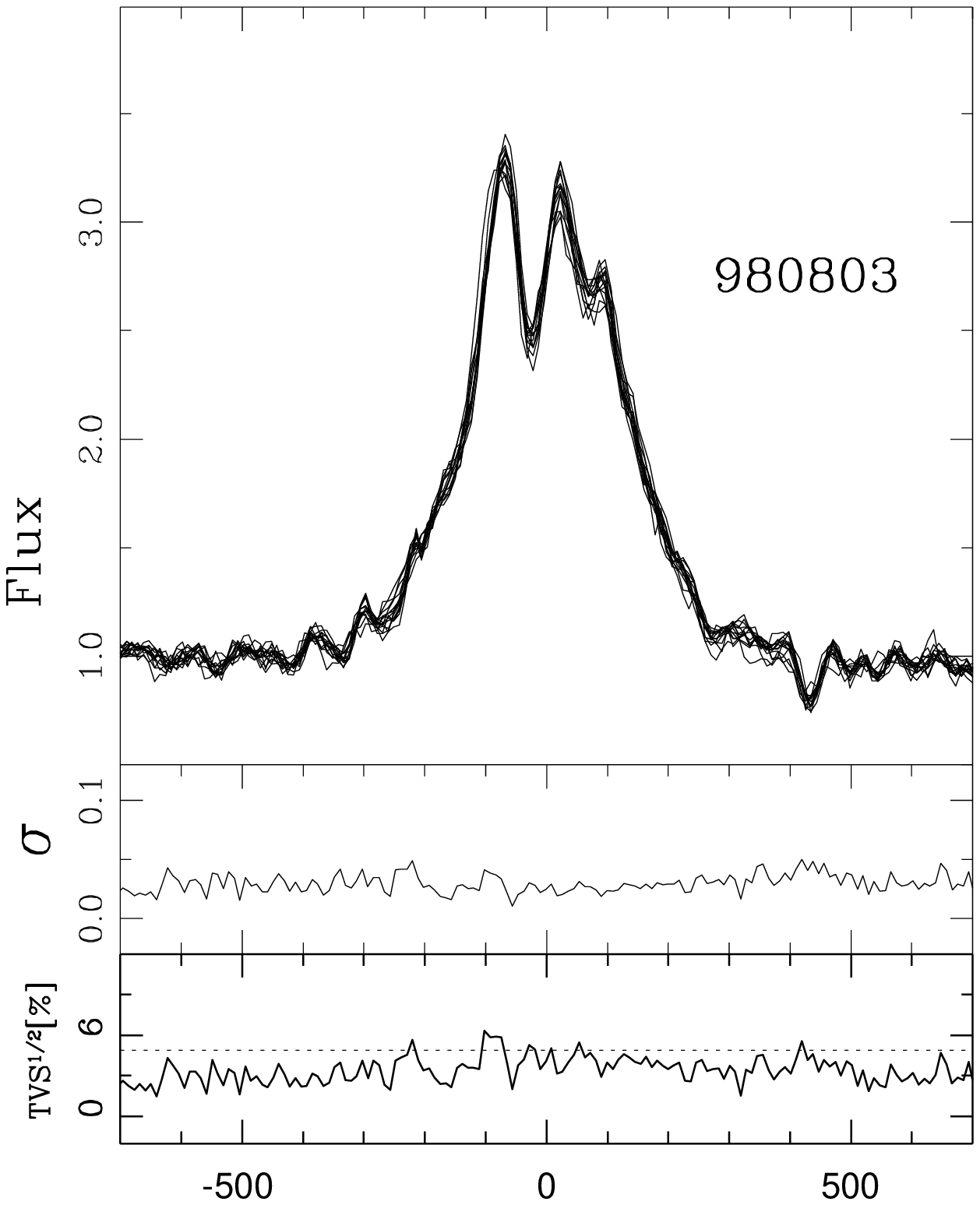}
  \includegraphics{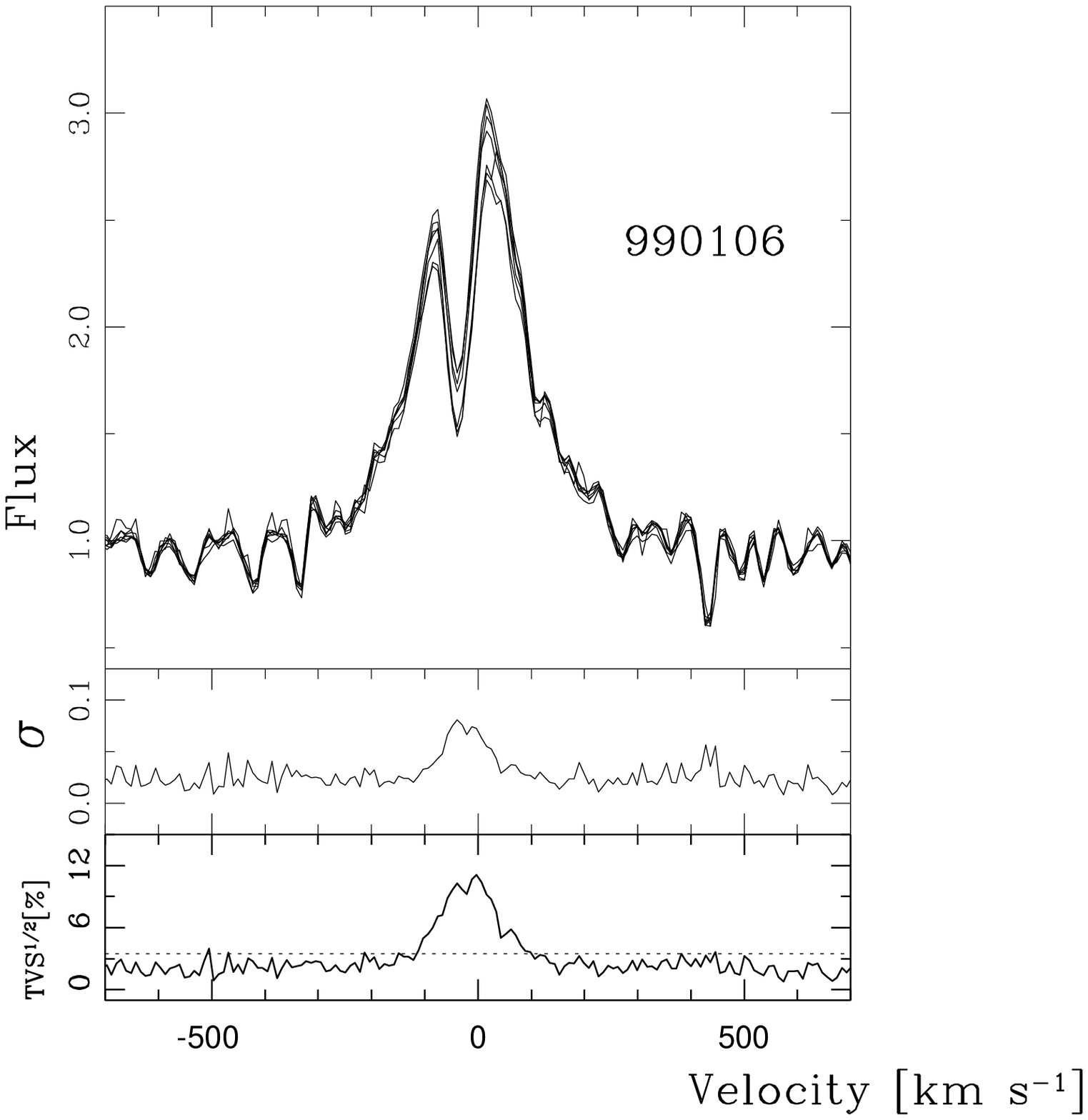}
  \includegraphics{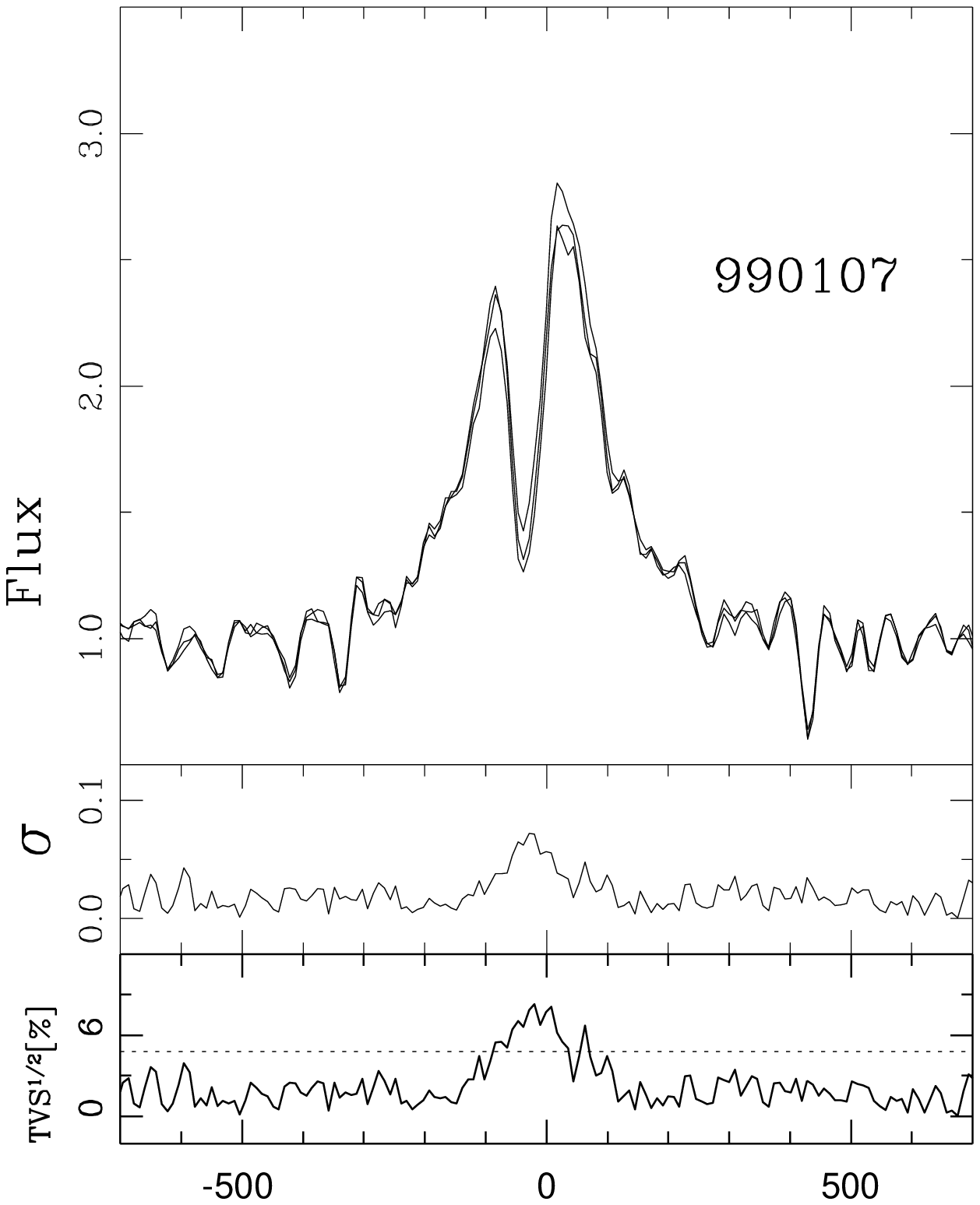}
\caption[]{ The \Ha\ profiles of T CrB for 5 nights of observations,
      normalised to the local continuum. The X-axis is in heliocentric velocity. 
      $\sigma$ and the TVS for all nights are plotted. The dotted horizontal line
      indicates a 1\% threshold for the detection of variability. 
      For first 3 nights $\sigma$
      does not indicate any variability, while the TVS is marginally above the
      1\% significance level on the night of 980415.
      On 990106 and 990107 well defined peaks, both in $\sigma$ and in the TVS, are
      clearly visible in the central part of the line, corresponding to 
      velocity  -100 to +100 km/s. This demonstrates that there is a 
      statistically significant variability in the profile of \Ha\ line with
      time resolution  $\sim$10 min. The probability that this is a real variability
      is $\ge$0.99.
   }
 \label{first}
\end{figure*}

The results of calculations are presented, 
together with all spectra, in Fig.\ref{first}.

As can be seen, the fractional variance spectra do not indicate variability in the first 
three nights (980415, 980802, and 980803), however the TVS indicates some variability 
in 980415. 
A clear peak  in both $\sigma$ and the TVS
is visible on 990106 and 990107 at about velocity $\approx$0 \kms. 

Following Fullerton, Gies \& Bolton (1996), if the TVS for a spectral 
feature is above a specified level of significance then the null hypothesis of
``no variability" can be rejected. The statistical distribution of 
the TVS is governed by the the reduced chi-squared distribution with N-1
degrees of freedom and scaled with SNR  (TVS$\sim s_0 ^2 \chi ^2 _{N-1}$).
For the nights considered here, the statistical 
significance of 1\% corresponds to TVS$^{1/2}$=3.6-5.0\%, 
depending on the number of spectra and SNR. 
In the central part of the line the TVS is  well above this level 
for 990106 and 990107 (Fig.\ref{first}). 
The peaks in $\sigma$ correspond practically to the same confidence level 
of 3 standard deviations from the mean.
This clearly shows that in these two nights 
there is statistically significant variability 
in the \Ha\ line profile, at a confidence level of 0.99, 
at a time resolution of $\sim$10 min.

\subsection{Variability in January 1999}
\begin{figure}
   \centering
   \includegraphics[width=10cm]{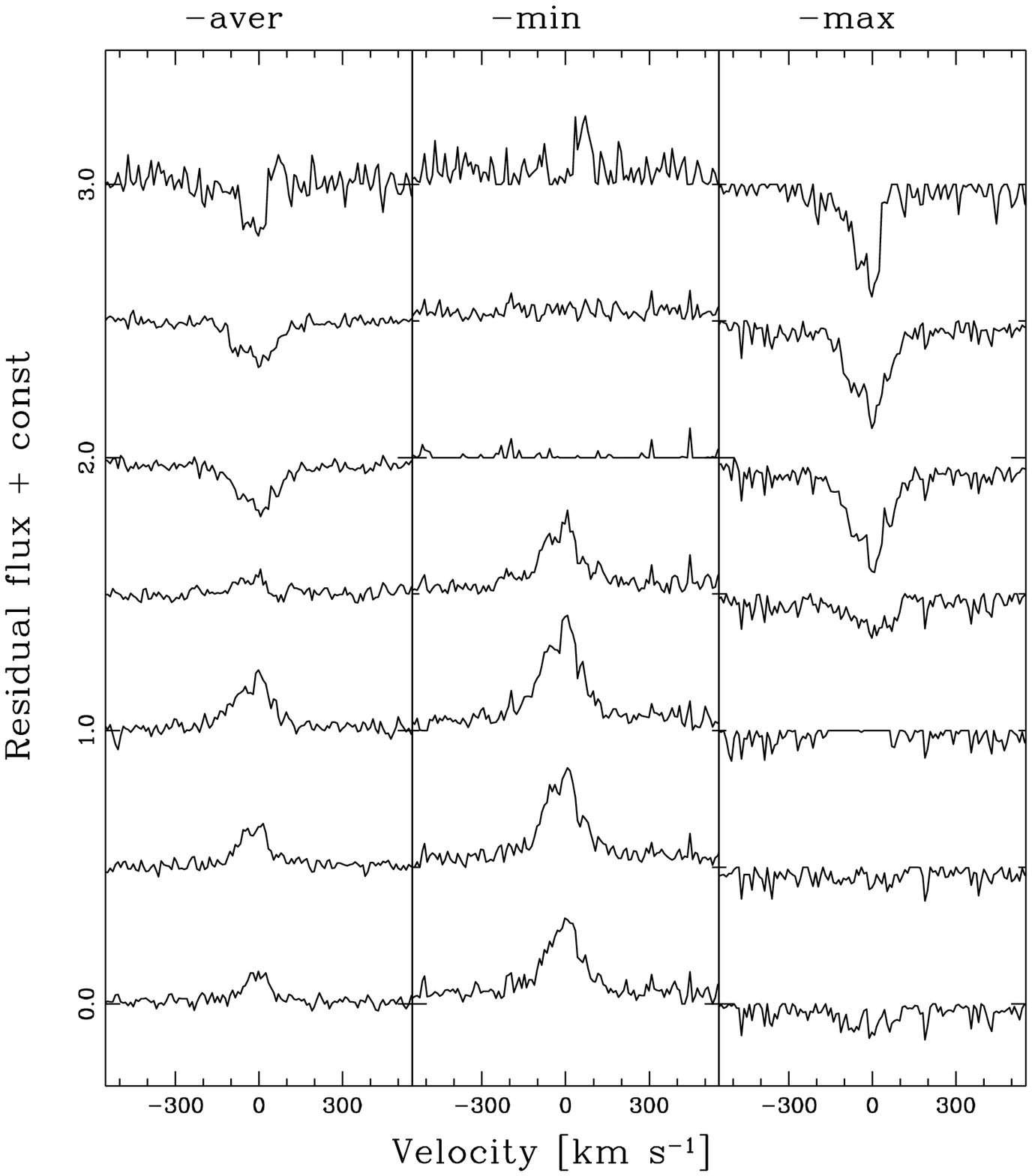}
      \caption{Night 990106: from the individual spectra have been subtracted 
      the  average spectrum (left panel), the minimum (middle panel), and 
      the maximum spectrum (right panel).
      Zero of the velocity axis  corresponds to the 
      systemic velocity  $\gamma$=-27.79 \kms (Fekel et al 2000).
      	      }
	 \label{dfourth}
\end{figure}

For the two nights with detected variability,
we subtract from the individual spectra the average spectrum, 
the minimum spectrum and the maximum spectrum.
The average spectrum is defined as the average of the normalised flux for every pixel.
The minimum   (maximum) spectrum is defined as the minimal (maximal) value for each pixel. 
We do not know {\it a priori} whether we have variability due to a pure 
emission, pure absorption, or an admixture of both.
The results for 990106 are presented in Fig. \ref{dfourth}.
As can be seen,  variability exists in all three panels. 
This indicates additional absorption/emission with equivalent width
of approximately 1 \AA, located  in the central part of the line. 

\begin{figure}
   \centering
   \includegraphics[width=9cm]{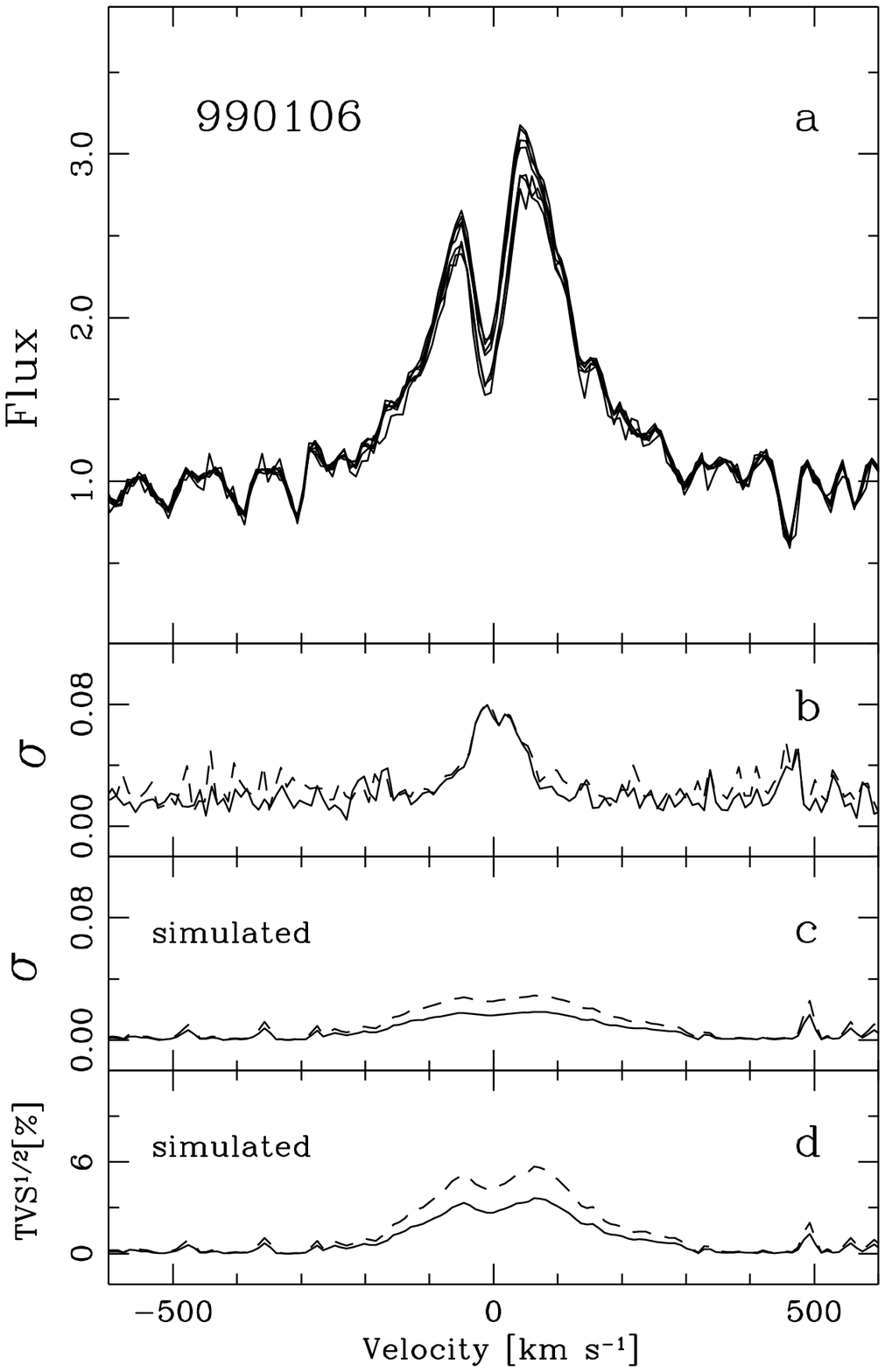}
      \caption{ 
      (a) \Ha\ profiles observed on 990106 (the same as in Fig.\ref{first}).
      (b) $\sigma$ for 990106, solid and dashed lines are for two deliberately 
      different extraction and normalisation procedures. 
      (c) simulated $\sigma$ as a result of hot continuum variability.
      (d) simulated TVS as a result of hot continuum variability.
      During the simulation we assumed that the hot continuum contributes 15\% 
      of the flux on average
      and it varies with $\pm$30\% (dashed line)  and $\pm$25\% (solid line).
      Zero of the velocity axis corresponds to the
      systemic velocity  $\gamma$.
            }
         \label{simul}
 \end{figure}

To exclude the possibility that the observed variability is an artifact
of data reduction
we performed tests with data processing and normalisation, which
showed that the detected variability on 990106 and 990107 is not an artificial 
result.
Fig.\ref{simul}b shows 
the results from two different extractions and
normalisations which have identical $\sigma$ peaks.

Although the sky conditions looked like photometric during the time of
the observations we can not exclude the possibility that the 
variable atmospheric lines and small wavelength shifts from 
spectrum to spectrum can result in spurious variability. 
To test the variability these effects could produce, we 
artificially introduced random shifts in wavelength up to 3 \kms
[the biggest shifts we detected were $<$0.25 px (2.5 \kms)].
We also  isolated athmosperic absorption lines from a high SNR spectrum 
of Spica ($\alpha$ Vir) and introduced variable 
atmospheric absorption in the spectra. We find that both efects can produce
variability, but this variability is lower than the noise in TVS and 
$\sigma$ and can not account for the observed peaks.

It is known (i.e. Zamanov \& Bruch 1998) that  the hot continuum,
which flickers on time scales of minutes 
with amplitude 0.2-0.3 mag in U, 
contributes about 15\% of the average flux in the V band.  
To check whether the variability detected in \Ha\ is due to 
the changing continuum level caused by 
the flickering of the hot component, we simulate $\sigma$ in such a case
by supposing that the hot continuum flickers with amplitude $\pm$25-30\%
and contributes 15\% of the flux in the continuum around \Ha. 
The amplitude is chosen to be slightly higher than the highest one 
observed in the U band after the subtraction of the red giant contribution
(see  Zamanov\& Bruch 1998).
To this variable 
hot continuum we add constant red giant and \Ha\ emission. The 
simulated $\sigma$ and TVS are shown in Fig.\ref{simul}. 
The simulated quantities are different from the ``observed" ones as (i) they have
smaller amplitude and (ii) they are wider and spread over the whole 
FWZI of the \Ha\ emission, 
while the observed ones are located in the central part of the
line only.  This indicates that detected variability 
is  not  due to the  continuum variability 
but is intrinsic to the line.

\section{Discussion}

The FWZI of the \Ha\ emission line is 800 \kms, with wings of the line 
formed in the immediate vicinity of the white dwarf 
(see also Stanishev et al. 2004).
However, the detected variability appears in the central part of the line and 
is confined to $\Delta V \approx \pm$100-120 \kms\ 
from the systemic velocity.
This indicates that this variability originates in the outer parts of the
\Ha\ emitting region.  Assuming Keplerian motion, a 1.4 M$_\odot$ white dwarf, 
and inclination 67$^0$ (i.e. Stanishev et al. 2004 and references therein)
this corresponds to a distance $\geq$20-25R$_\odot$.

Mass accretion in T CrB is occurring via Roche lobe overflow. 
As the stream of gas flows away from L$_1$ it forms a ring 
(see Verbunt 1982). The position of the  ring 
approximately defines the outer edge of the accretion disk. 
The radius of the ring can be estimated from (Hessman \& Hopp 1990):
$r_r/a=0.0859(q)^{-0.426}$, where $a$ is the semi-major axis and $q$
is the mass ratio of giant to white dwarf.
For T CrB  q=0.82$\pm$0.10 
and $a \approx 210$ R$_\odot$ (Stanishev et al. 2004), which places the ring 
at 20 R$_\odot$. At this location, the Keplerian velocity is about 
115 \kms, which is similar to  $\Delta V$ of the detected variability. 
However, the Keplerian time scale is about 18 days, 
which is not comparable with the timing of our observations. This
suggests that the variability is not produced by an inhomogeneous disc
structure.

The observed behaviour of \Ha\ emission from T~CrB with time resolution 10-15 minutes
is different from that of CH~Cyg and of the variable winds observed in some CVs. 
We do not see ejected blobs, like those observed in CH~Cyg (by Tomov et
al. 1996), nor high velocity 
components observed in the variable disk winds of some CVs
(Prinja et al. 2003).
It might be that they are not detectable in our H$\alpha$ profiles, even
though they may exist and could be detected in the UV, where the hot component
is the dominant source of radiation.

One possible explanation for the detected variability could be
variable absorption from the wind of the giant or from the
accretion stream. It is worth noting, that 
no variability in the line is observed
at orbital phases where the line of sight passes through the expected position 
of maximum absorption from the red giant wind.  
Another cause could be additional emission from the area where the accretion stream 
hits the disk (the hot spot).
It is possible that a variable component with EW$\le$0.5\AA\
always exists, but when the line is strong it is lost in the line
(e.g. 980415).

An additional plausible explanation is that  
the variable continuum arising from the central part of the accretion disk 
(where the flickering in U arises) changes the ionisation state of the
outer regions of the (flared) disk, from which the central part of the
emission lines arise. This possibility deserves further exploration.

\section{Conclusions}

We searched for rapid \Ha\ variability in 
the spectra of the recurrent nova T~CrB. 
On 2 nights out of 5, we detect statistically significant variability 
at time resolution of 10-20 minutes. 
The detected variability is confined to the central part 
of the line profile at $\pm$100 \kms, which suggests
that variations are produced in the outer parts of the accretion disk 
($>$20 R$_\odot$).
On these two nights there is evidence for 
variability of the total EW(\Ha) (in all cases it is calculated relative 
to the average value, see Table 1)
with amplitude $\pm$8\% for 990106 and $\pm$6\%  for 990107,
which is approximately five times the error of 
the individual measurements.

For the other 3 nights of observations we do not detect line profile changes. 
We place upper limits on the variability of the total EW(\Ha) 
$\Delta$EW(\Ha) $\pm$2\% for 980415,  
$\pm$4\% for 980802 and $\pm$3\% for 980803.

More extensive spectral observations with better time resolution, 
combined with photometry  
are necessary to  define what influences the \Ha\ variability and 
whether this variability is directly connected with the 
flickering.


\end{document}